\documentclass[referee,sn-nature]{sn-jnl}

\usepackage{graphicx}%
\usepackage{multirow}%
\usepackage{amsmath,amssymb,amsfonts}%
\usepackage{amsthm}%
\usepackage{mathrsfs}%
\usepackage[title]{appendix}%
\usepackage{xcolor}%
\usepackage{textcomp}%
\usepackage{manyfoot}%
\usepackage{booktabs}%
\usepackage{array}
\usepackage{algorithm}%
\usepackage{algorithmicx}%
\usepackage{algpseudocode}%
\usepackage{listings}%
\usepackage{siunitx}
\usepackage{chemformula}
\usepackage{subcaption}
\usepackage{setspace}
\usepackage{graphicx}
\usepackage{adjustbox}

\makeatletter
\providecommand\add@text{}
\newcommand\tagaddtext[1]{%
	\gdef\add@text{#1\gdef\add@text{}}}%
\renewcommand\tagform@[1]{%
	\maketag@@@{\llap{\add@text\quad}(\ignorespaces#1\unskip\@@italiccorr)}%
}
\makeatother

\makeatletter
\newcommand{\mathleft}{\@fleqntrue\@mathmargin0pt}
\newcommand{\mathcenter}{\@fleqnfalse}
\makeatother

\theoremstyle{thmstyleone}%
\theoremstyle{thmstyletwo}%

\theoremstyle{thmstylethree}%

\begin{document}

\title[Article Title]{Speciation controls the kinetics of iron hydroxide precipitation and transformation}


\author[1]{\fnm{Fabio E.} \sur{Furcas}}\email{ffurcas@ethz.ch}

\author[1]{\fnm{Shishir} \sur{Mundra}}\email{smundra@ethz.ch}
\equalcont{These authors contributed equally to this work.}

\author[2]{\fnm{Barbara} \sur{Lothenbach}}\email{barbara.lothenbach@empa.ch}
\equalcont{These authors contributed equally to this work.}

\author*[1]{\fnm{Ueli M.} \sur{Angst}}\email{uangst@ethz.ch}
\equalcont{These authors contributed equally to this work.}

\affil[1]{\orgdiv{Institute for Building Materials}, \orgname{ETH Z\"{u}rich}, \orgaddress{\street{Laura-Hezner-Weg 7}, \city{Z\"{u}rich}, \postcode{8093}, \country{Switzerland}}}

\affil[2]{\orgdiv{Concrete \& Asphalt Laboratory}, \orgname{Empa}, \orgaddress{\street{Ueberlandstrasse 129}, \city{D\"{u}bendorf}, \postcode{8600}, \country{Switzerland}}}


\abstract{The formation of energetically favourable and metastable mineral phases within the \ch{Fe}$-$\ch{H2O} system controls the long-term mobility of iron complexes as well as other aqueous phase constituents in natural aquifers, soils and other environmentally and industrially relevant systems. The fundamental mechanism controlling the formation of these solid phases has remained enigmatic. Here, we develop a general state-of-the-art partial equilibrium model and succeed in predicting the rate of amorphous 2-line ferrihydrite precipitation, dissolution and overall transformation to crystalline goethite at alkaline pH. All mechanistic steps constituting the transformation mechanism accurately predict the experimentally measured solid and aqueous phase composition over time, involving only a single kinetic rate constant each.  It is found that the precipitation of goethite (i) occurs from solution and (ii) is limited by the comparatively slow dissolution of the first forming amorphous phase 2-line ferrihydrite. A generalised transformation mechanism applicable to near-neutral and mildly acidic pH further illustrates that differences in the kinetics of Fe(III) precipitation are controlled by the coordination environment of the predominant Fe(III) hydrolysis product. Findings provide a framework for the modelling of other iron(bearing) phases across a broad range of aqueous phase compositions.  
}




\maketitle
\newpage
\clearpage
\section*{Main}\label{sec:Main}
Depending on the aqueous phase composition and a range of other physiochemical parameters including temperature and the pH, iron may precipitate in the form of over 38 stable and metastable phases characterised to date \cite{lemire2013chemical}. 
Iron (hydr)oxides are the most common form of metallic oxides in soils \cite{schwertmann_taylor_iron_oxides}. Their formation governs the immobilisation of elements of concern (EOCs) including \ch{As}, \ch{Se}, \ch{Mo}, \ch{Ni} and \textsuperscript{226}\ch{Ra} in groundwater streams \cite{RN237}, soil environments \cite{RN253}, nuclear processing facilities \cite{RN229} and across a broad range of other natural and industrial aqueous systems \cite{RN235,RN236,RN259}. Iron (hydr)oxide precipitation within the pore network of cementitious materials is one of the major causes for premature structural degradation of reinforced concrete structures \cite{angst2018challenges}. Iron uptake by calcium silicate hydrates (C-S-H) may further reduce the ability of cement-based nuclear waste repositories to contain and safely store hazardous radionuclides  \cite{wieland_strontium,tits_update}. Iron (hydr)oxides are also versatile industrial products used as pigments in the production of paints and coatings \cite{RN262}, in wastewater treatment \cite{RN265}, as well as in nanotechnology \cite{RN254}, photovoltaic \cite{RN250} and energy storage systems \cite{RN263}. For these reasons, detailed knowledge about the mechanism and transformation kinetics of such iron (hydr)oxides is needed to assess their stability over different time scales and conditions. \\

\begin{figure}[!ht]
	\centering
	\includegraphics[scale=0.65]{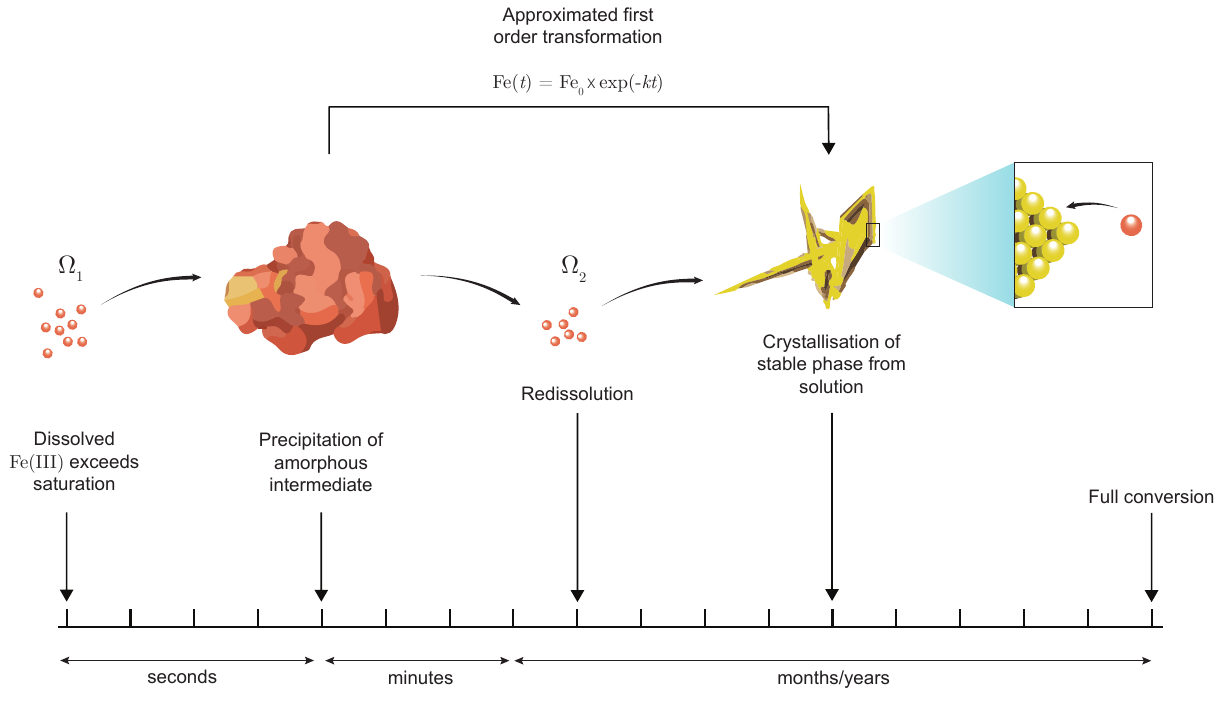}
	\caption{Schematic representation of the formation mechanism of crystalline iron (hydr)oxide phases from supersaturated aqueous solutions. The initial precipitation of amorphous intermediate compounds from dissolved Fe(III) proceeds within seconds, full conversion to the stable end member is reached after months to years. This multi-step conversion process is commonly approximated by the integrated first order rate equation $\text{Fe}(t) = \text{Fe}_0\times \text{exp}(-kt)$, irrespective of the varying degree of supersaturation $\Omega$ and other aqueous phase parameters.}
	\label{fig:schematic}
\end{figure}

\noindent Investigations into the kinetics of iron (hydr)oxide precipitation primarily quantify the reaction rate, and thus its extent, by monitoring the molar fraction of solids formed, often assuming direct proportionality between their rate of formation and concentration \cite{schwertmann_ph_ferri,das_arsenate,twoline_gt_transformation,RN267}. Two objections may be raised against this modelling approach. Firstly, it is known that the formation and transformation of some iron (hydr)oxides proceeds via particle-mediated growth mechanisms \cite{soltis2016phase,RN184}, or involve metastable intermediate species \cite{twoline_gt_transformation} (Fig. \ref{fig:schematic}). As opposed to growth by the addition of singular atoms into an existing solid phase, as described within framework of classical nucleation theory \cite{baumgartner2013nucleation}, these nonclassical growth mechanisms consist of multiple dissolution and precipitation steps and can thus not be described completely by the integrated first order rate equation or any other semi-empirical equation of the form $\text{Fe}(t) = \text{Fe}_0\times \text{exp}(f(t))$. 
Secondly, mineral dissolution and growth rates are, amongst other parameters, dependent on the aqueous phase composition, the degree of supersaturation $\Omega$ and the activity of dissolved species in disequilibrium with the solid phase(s). These parameters are generally not considered in first order rate expressions. Moreover, due to the low solubility of iron, these parameters are significantly more difficult to obtain from an experimental point of view than the molar fraction of solid phases. 

\noindent In the simplest case, the irreversible formation of one ferrous ($\text{z}=2$) or ferric ($\text{z}=3$) iron (hydr)oxide according to the general reaction
\begin{equation}
	\ch{Fe^{z+}} + (\text{r}+\text{s})\ch{H2O(l)} \rightarrow \ch{FeO_r(OH)_s(s)^{\text{z}-2\text{r}-\text{s}}} + (2\text{r}+\text{s})\ch{H^{+}}
\end{equation}
is expected to depend on the \ch{Fe^{z+}} and the \ch{H^+} activity. In the context of natural and industrially relevant aqueous electrolytes, the phase assemblage of iron (hydr)oxides is significantly more complex. Consider the fate of \ch{Fe^{2+}} due to the corrosion of carbon steel in near-neutral environments. In the aqueous phase, the ferrous cation may be coordinated as \ch{FeOH^+}, \ch{Fe(OH)2(aq)} or \ch{Fe(OH)3^-}, depending on the pH. These aqueous Fe(II) complexes may further oxidise, both aerobically and in the absence of oxygen, to form \ch{Fe^{3+}} or any of the Fe(III) hydrolysis products \ch{FeOH^{2+}}, \ch{Fe(OH)2^+}, \ch{Fe(OH)3(aq)} or \ch{Fe(OH)4^-} \cite{mundra_oxidation}. The presence of carbonates, chlorides and silica or any other anion characteristic to the aqueous environment of interest \cite{bottero_iron_hydroxide_chloride,ferrihydrite_silicates,RN227,RN229} can lead to further complexation of the dissolved Fe(II) and Fe(III) hydrolysis products. The phase assemblage of solid iron(bearing) phases is thus in direct competition with the speciation of iron in the aqueous phase. From all these considerations, it is evident that the mechanism fundamentally controlling the kinetics of iron (hydr)oxide formation can only be unravelled in a tightly coupled investigation of both, the solid and the aqueous phase composition. Until recently, however, there was no data reported that comprehensively characterises the evolution of both the solid phases and the electrolyte composition. \\
Recent studies \cite{furcas_transformation_est,pham_precipitation} reporting on the time evolution of solid Fe(III) hydroxides and complexes allow, for the first time, to model their formation mechanism in the $\ch{Fe}-\ch{H2O}$ system under full consideration of the aqueous phase in disequilibrium with one or more of these minerals. On this basis, we develop a new partial equilibrium model, combining state-of-the-art thermodynamic parameters and particle morphology-dependent kinetic rate equations. We use this model to demonstrate that formation of goethite, a thermodynamically stable iron hydroxide, is controlled by the dissolution kinetics of amorphous 2-line ferrihydrite at alkaline pH. All elementary steps constituting the dissolution-crystallisation pathway rely on a single kinetic rate constant. Upon considering the speciation of aqueous Fe(III) at neutral to mildly acidic pH, the transformation mechanism can be generalised to all thermodynamically stable solid Fe(III) phases. Here, the overall rate of Fe(III) precipitation is limited by the intrinsic precipitation rate of the predominant hydrolysed aqueous Fe(III) species, \ch{Fe(OH)3(aq)} at circumneutral and \ch{Fe(OH)4^-} at alkaline pH. These observations are in line with both Stranski's Rule \cite{blesa_stranski} and the Ostwald Step Rule \cite{ostwald1897studien}. We envision this model to be expanded to a wider range of iron-bearing phases and aqueous systems.	

\subsection*{The precipitation of 2-line ferrihydrite at alkaline pH}
\begin{figure}[!ht]
	\centering
	\includegraphics[scale=0.9]{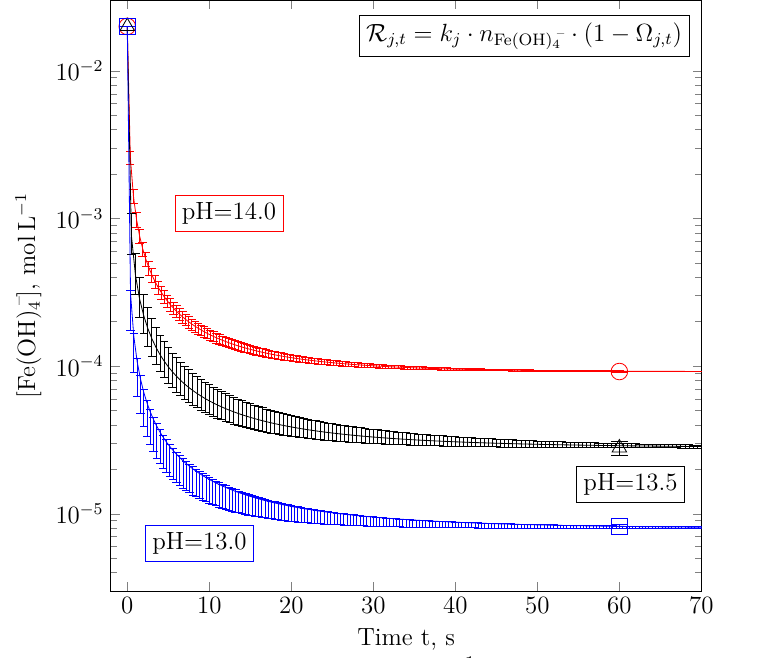}
	\caption{The modelled progression of aqueous $[\ch{Fe(OH)4^-}]$ (solid lines) compared to the experimentally measured aqueous iron concentration (symbols) in supersaturation with respect to 2-line ferrihydrite, as determined by Furcas et al. \cite{furcas_transformation_est} within the first 60 seconds of equilibration time. Error bars represent the standard deviation of the simulated concentration profiles fitted to the upper and lower experimentally measured aqueous iron concentration. The estimated apparent rate constant of transformation $(k_{j}\cdot A_{j,t}) = 0.078 \pm 0.010$ \si{\per\second} and reaction order $w_j = 1$ are independent of the pH.}
	\label{fig:2l_precip_errorbar}
\end{figure}

Recently, we showed that the precipitation of 2-line ferrihydrite (2l$-\ch{Fe(OH)3(s)}$) from supersaturated alkaline stock solutions (e.g. $[\ch{Fe(III)}]>10^{-4}$ M at pH = 14.0) occurs significantly more rapidly, than its transformation to more stable secondary phases including hematite ($\alpha-\ch{Fe2O3(s)}$) and goethite ($\alpha-\ch{FeOOH(s)}$) \cite{furcas_transformation_est}. As over 99.8 $\%$ of Fe(III) in solution is coordinated as \ch{Fe(OH)4^-} at a pH $\geq 12$ \cite{furcas_solubility}, the precipitation of $2\text{l}-\ch{Fe(OH)3(s)}$ at alkaline pH can be described by:
\begin{equation} \label{eq:2lprecip}
	\ch{Fe(OH)4^-} + \ch{H^+} \rightarrow 2\text{l}-\ch{Fe(OH)3(s)} + \ch{H2O(l)}.
\end{equation}
Considering that phase growth velocity is anticipated to rise with increasing activity of $\ch{Fe(OH)4^-}$ and decline as saturation conditions are approached, we formulate the rate of 2-line ferrihydrite precipitation as
\begin{equation}
	\mathcal{R}_{j,t} = k_{j}\cdot n_{\ch{Fe(OH)4^-},t}^{w_j} \cdot(1 - \Omega_{j,t}), \qquad j=2\text{l}-\ch{Fe(OH)3(s)}, \ \forall t.
\end{equation}
Correspondingly, the molar balance\footnote{The stoichiometric coefficients of all species in Reaction \ref{eq:2lprecip} are $\nu_i=1$. They are omitted from the molar balance displayed in Equation \ref{eq:2lprecip_molar_balance}.} of all species $i$ involved in the formation reaction displayed in Equation \ref{eq:2lprecip} are
\begin{equation} \label{eq:2lprecip_molar_balance}
	\frac{\partial n_i}{\partial t} = \nu_iA_{j,t}\mathcal{R}_{j,t} = (A_{j,t}\cdot k_j)\cdot n_{\ch{Fe(OH)4^-},t}^{w_j} \cdot(1 - \Omega_{j,t}).
\end{equation}
\noindent To quantify the rate constant $k_j$ and reaction order $w_j$ as a function of the pH, the progression of $[\ch{Fe(OH)4^-}]$ is fitted to the aqueous Fe(III) concentration, measured by inductively coupled plasma optical emission spectroscopy (ICP-OES) at pH = 13.0, 13.5, 14.0, within the first 60 seconds of equilibration time. It is assumed that the morphology of precipitated 2-line ferrihydrite does not change upon growth. Fig. \ref{fig:2l_precip_errorbar} displays the resultant concentration profiles at various pH over time. It can be recognised that the precipitation rate drastically decreases, as the aqueous \ch{Fe(OH)4^-} concentration approaches its pH-dependent solubility limit with respect to 2-line ferrihydrite.
Within the pH interval investigated, the solubility limit of 2-line ferrihydrite increases by approximately one order of magnitude per pH unit. The pH dependence of $\mathcal{R}_{j,t}$ is therefore implicitly accounted for by the thermodynamic speciation solver. Within the error of the experimentally measured Fe(III) concentrations, the apparent rate constant of 2l$-\ch{Fe(OH)3(s)}$ precipitation is evaluated as $(k_{j}\cdot A_{j,t}) = 0.078 \pm 0.010$ \si{\per\second}, whilst the reaction order is $w_j = 1$ with respect to the \ch{Fe(OH)4^-} concentration. 
\newpage
\clearpage
\subsection*{The transformation of 2-line ferrihydrite to goethite at alkaline pH} \label{sec:transformation_ferri_gt}
\begin{figure}[!ht]
	\centering
	\hspace{-25pt}
	\begin{subfigure}[b]{\textwidth}
		\centering
		\includegraphics[scale=0.9]{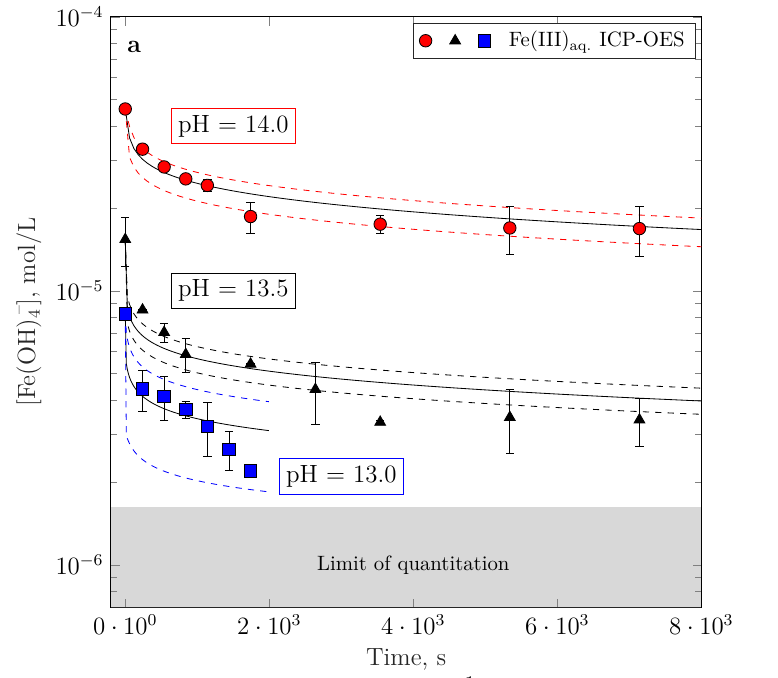}
	\end{subfigure}%
	\vskip\baselineskip
	\begin{subfigure}[b]{\textwidth}
		\centering
		\includegraphics[scale=0.9]{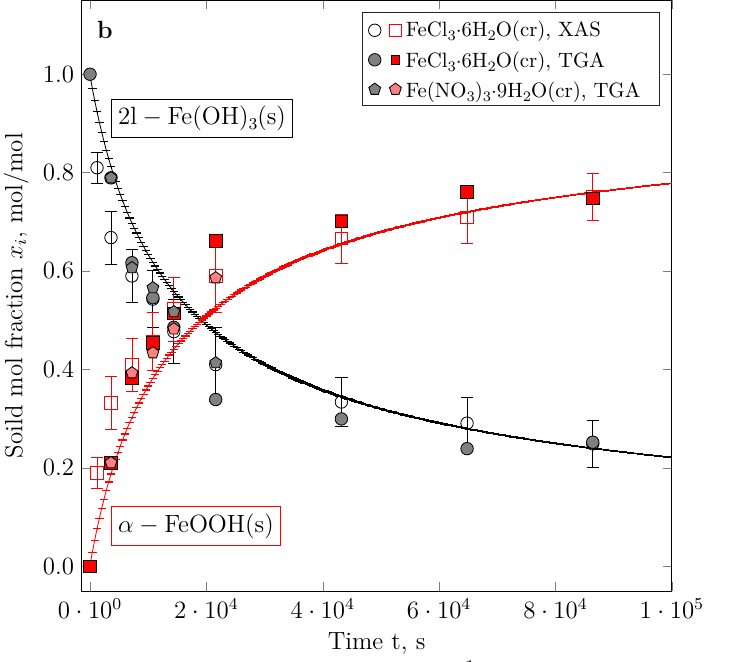}
	\end{subfigure}
	\caption{Combined modelling results (solid lines) obtained by adjusting the kinetic parameters governing the rate of goethite formation (Equation \ref{eq:rate_gt_precip}) and 2-line ferrihydrite dissolution (Equation \ref{eq:rate_2l_diss}) to match the iron concentration measured by ICP-OES (Fig. \ref{fig:modelling_all}a) and the molar fraction of both minerals as determined by XAS and TGA formed from $\ch{FeCl3}\cdot6\ch{H2O(cr)}$ and $\ch{Fe(NO3)3}\cdot9\ch{H2O(cr)}$ at pH = 14.0 (Fig. \ref{fig:modelling_all}b), measured by Furcas et al. \cite{furcas_transformation_est}. Dashed lines represent the confidence interval of the computationally predicted aqueous iron concentrations.  Simulations correct for the specific surface area of both phases by applying a mass cubic root correction, as displayed in Equation \ref{eq:mass_cubic_root}.}
	\label{fig:modelling_all}
\end{figure}

\begin{figure}[!ht]
	\centering
	\begin{subfigure}[b]{\textwidth}
		\centering
		\includegraphics[scale=0.9]{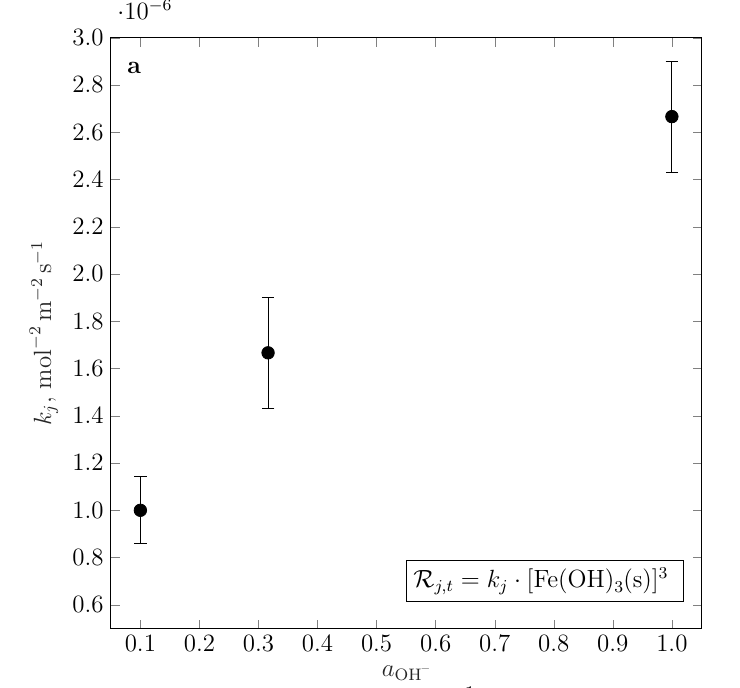}
	\end{subfigure}%
	\vskip\baselineskip
	\begin{subfigure}[b]{\textwidth}
		\centering
		\includegraphics[scale=0.9]{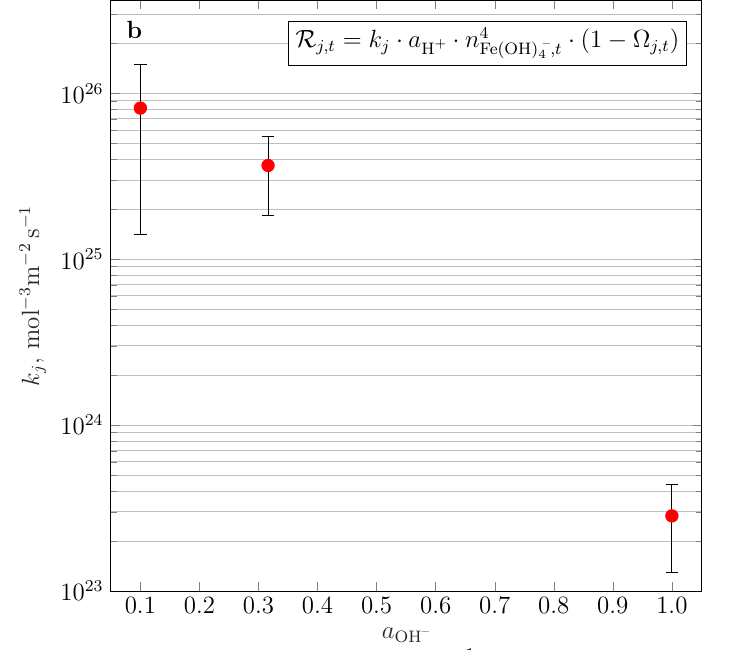}
	\end{subfigure}
	\caption{Estimated surface rate constants $k_j$ in mol$^{-3}$\si{\per\square\meter\per\second} of 2-line ferrihydrite dissolution (Fig. \ref{fig:rate_constants_all}a) and goethite precipitation (Fig. \ref{fig:rate_constants_all}b) as a function of the pH.}
	\label{fig:rate_constants_all}
\end{figure}

\begin{figure}[!ht]
	\centering
	\hspace{-25pt}
	\begin{subfigure}[b]{\textwidth}
		\centering
		\includegraphics[scale=0.9]{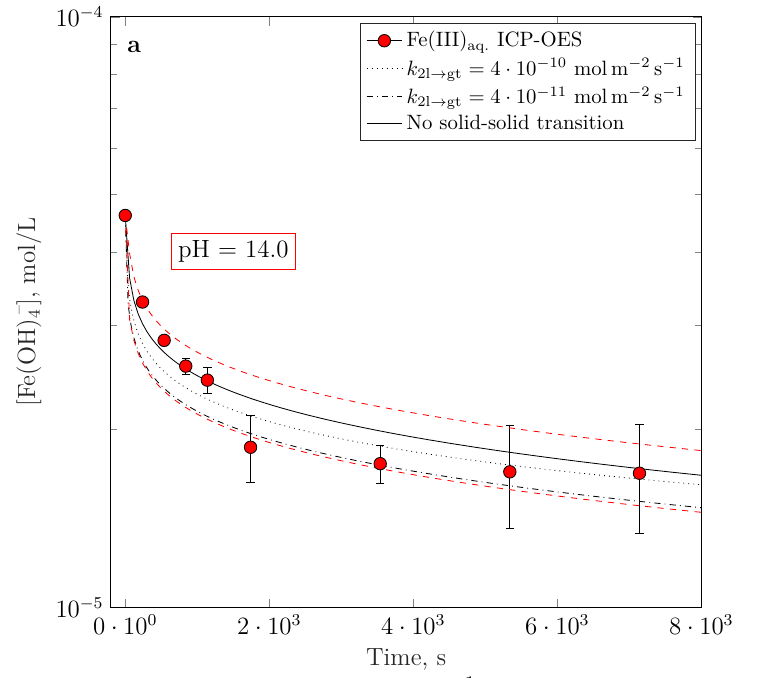}
	\end{subfigure}%
	\vskip\baselineskip
	\begin{subfigure}[b]{\textwidth}
		\centering
		\includegraphics[scale=0.9]{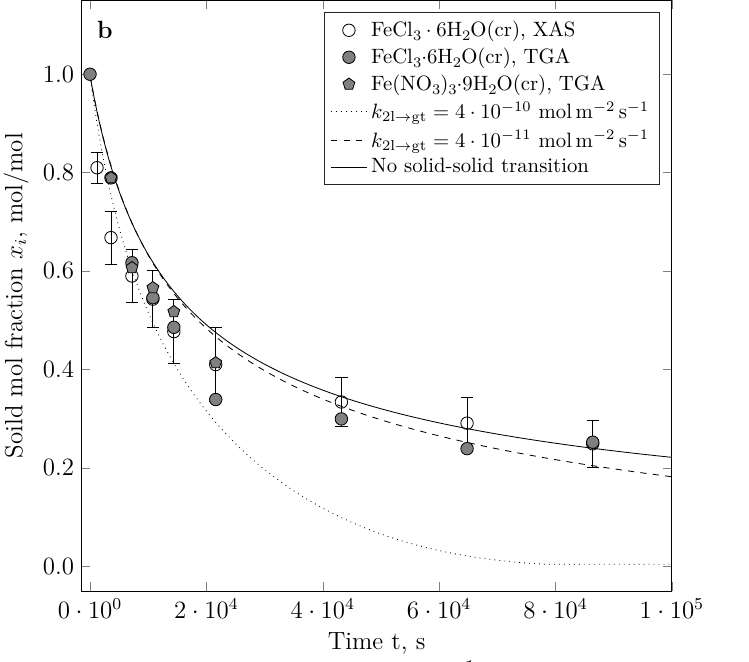}
	\end{subfigure}
	\caption{The predicted aqueous \ch{Fe(OH)4^-} concentration (Fig.\ref{fig:modelling_solid_solid}a) and solid molar fraction of 2-line ferrihydrite (Fig \ref{fig:modelling_solid_solid}b) including a solid-solid transition reaction between 2-line ferrihydrite and goethite at pH = 14.0. In both figures, the dashed and dotted lines correspond to the simulation results including a low ($k_{2\text{l}\rightarrow\text{gt}} = 4\cdot10^{-11}$ \si{\mole\per\square\meter\per\second}) and high ($k_{2\text{l}\rightarrow\text{gt}} = 4\cdot10^{-10}$ \si{\mole\per\square\meter\per\second}) rate of solid-solid transformation.}
	\label{fig:modelling_solid_solid}
\end{figure}

\begin{figure}[!ht]
	\centering
	\begin{subfigure}[b]{\textwidth}
		\centering
		\includegraphics[scale=0.9]{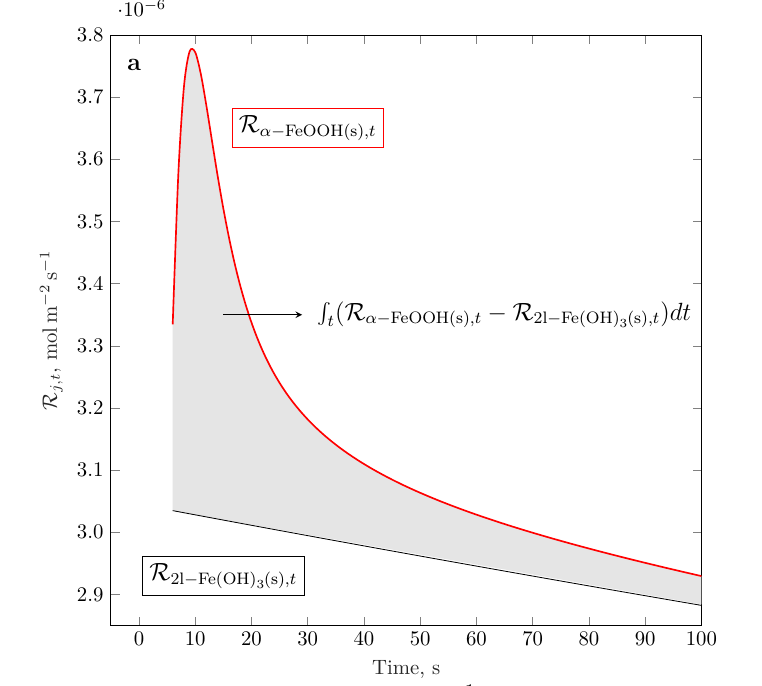}
	\end{subfigure}%
	\vskip\baselineskip
	\begin{subfigure}[b]{\textwidth}
		\centering
		\includegraphics[scale=0.9]{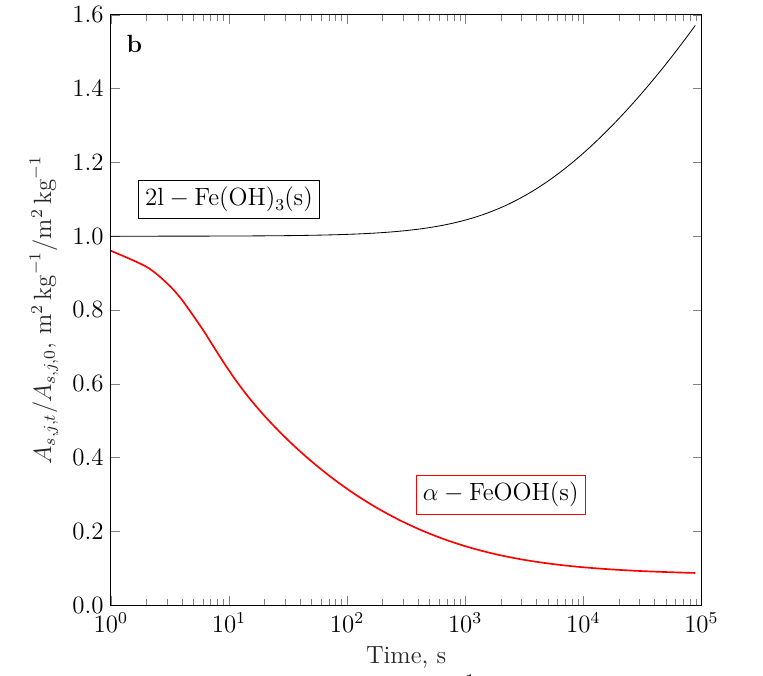}
	\end{subfigure}
	\caption{Rates of 2-line ferrihydrite dissolution an goethite precipitation $\mathcal{R}_{j,t}$ in \si{\mol\per\square\meter\per\second} (Fig. \ref{fig:rates_surfaces}a) and their normalised specific surface areas $A_{s,j,t}/A_{s,j,0}$ in \si{\square\meter\per\kilo\gram}/\si{\square\meter\per\kilo\gram} (Fig. \ref{fig:rates_surfaces}b) at pH = 14.0 over time.}
	\label{fig:rates_surfaces}
\end{figure}

For aqueous Fe(III) concentrations in-between the solubility limits of 2-line ferrihydrite and goethite (e.g. $10^{-4} > [\ch{Fe(III)}]>10^{-7}$ M at pH = 14.0), the aqueous \ch{Fe(OH)4^-} concentration can increase due to the re-dissolution of 2l$-$\ch{Fe(OH)3(s)} according to
\begin{equation} \label{eq:2ldiss}
	2\text{l}-\ch{Fe(OH)3(s)} + \ch{H2O(l)} \rightarrow \ch{Fe(OH)4^-} + \ch{H^+}
\end{equation} 
and decrease due to the precipitation of goethite from solution
\begin{equation} \label{eq:gtprecip}
	\ch{Fe(OH)4^-} + \ch{H^+} \rightarrow \alpha-\ch{FeOOH(s)} + 2\ch{H2O(l)}.
\end{equation}
Phase growth may also occur via aggregation-based mechanisms, involving the formation of iron-oxygen bonds due to the elimination of water:
\begin{equation} \label{eq:solid_solid}
		2\text{l}-\ch{Fe(OH)3(s)} \rightarrow \alpha-\ch{FeOOH(s)} + \ch{H2O(l)}.
\end{equation}
Analogous to the formation of 2-line ferrihydrite ($2\text{l}$), the growth rate of goethite ($\text{gt}$) is expected to be primarily dependent on the activity of \ch{Fe(OH)4^-} as well as the degree of supersaturation
\begin{equation} \label{eq:rate_gt_precip}
	\mathcal{R}_{\text{gt},t} = k_{j}\cdot n_{\ch{Fe(OH)4^-},t}^{w_{j}} \cdot(1 - \Omega_{j,t}), \qquad j=\alpha-\ch{FeOOH(s)}, \ \forall t.
\end{equation}
In contrast, the rate of 2-line ferrihydrite dissolution 
\begin{equation} \label{eq:rate_2l_diss}
	\mathcal{R}_{2\text{l},t} = k_{j}\cdot n_{j,t}^{w_{j}}\cdot(1-\Omega_{j,t}), \qquad j=2\text{l}-\ch{Fe(OH)3(s)}, \ \forall t
\end{equation} 
is found to be insensitive to the saturation index $\Omega$, as the aqueous $\ch{Fe(OH)4^-}$ concentration remains close to the solubility limit of 2-line ferrihydrite (Supporting information, Fig. \ref{fig:saturation_indices}). 
It is instead determined by the number of moles of $n_{2\text{l},t}$. The rate of aggregation-based growth of goethite from 2-line ferrihydrite does not involve the redissolution of \ch{Fe(OH)4^-}, and is thus written as 
\begin{equation} \label{eq:rate_solid_solid}
	\mathcal{R}_{2\text{l}\rightarrow\text{gt}} = k_j \cdot n_{j,t}^{w_{j}}, \qquad j=2\text{l}-\ch{Fe(OH)3(s)},  \ \forall t.
\end{equation}
Combining these rate expressions\footnote{Apart from the stoichiometric coefficient of \ch{H2O(l)} in the precipitation reaction of goethite, various other $\nu_i$ in Reactions \ref{eq:2ldiss}, \ref{eq:gtprecip} and \ref{eq:solid_solid} equal to 1. They are omitted from the combined molar balance displayed in Equation \ref{eq:gtprecip_molar_balance}.}, the molar species balances that describe the evolution of various species $i$ involved in the dissolution (Equation \ref{eq:2ldiss}) and precipitation (Equation \ref{eq:gtprecip}) reaction are
\begin{align}
\frac{\partial n_i}{\partial t} &= \nu_{i,\text{gt}}A_{\text{gt},t}\mathcal{R}_{\text{gt},t} + \nu_{i,2\text{l}}A_{2\text{l},t}\mathcal{R}_{2\text{l},t} + \nu_{i,2\text{l}\rightarrow\text{gt}}A_{2\text{l},t}\mathcal{R}_{2\text{l}\rightarrow\text{gt}} \nonumber \\
&= (A_{\text{gt},t} \cdot k_{\text{gt}}) \cdot n_{\ch{Fe(OH)4^-},t}^{w_{\text{gt}}}\cdot(1-\Omega_{\text{gt},t}) + (A_{2\text{l},t}\cdot k_{2\text{l}})\cdot n_{2\text{l},t}^{w_{2\text{l}}} \nonumber \\
&+ (A_{2\text{l},t}\cdot k_{2\text{l}\rightarrow\text{gt}})\cdot n_{2\text{l},t}^{w_{2\text{l}\rightarrow\text{gt}}} \label{eq:gtprecip_molar_balance}
\end{align}
\noindent Taking the initial specific surface area of 2-line ferrihydrite and goethite to be $6.0\cdot 10^5$ and $1.3\cdot 10^5$ \si{\square\meter\per\kilo\gram} \cite{schwertmann2008iron}, the kinetic rate parameters of $\partial n_i/\partial t$ are determined by fitting the predicted aqueous progression of $[\ch{Fe(OH)4^-}]$ and the solid mole fraction $x_j$ of both phases $j\in\Gamma$ to the experimental data collected by Furcas et al. \cite{furcas_transformation_est}. It is assumed that the specific surface area $A_{s,j}$ scales with the phase mass in accordance the cubic root correction formula displayed in Equation \ref{eq:mass_cubic_root}. As illustrated in Fig. \ref{fig:modelling_all}, the calculated aqueous Fe(III) concentration (Fig. \ref{fig:modelling_all}a) and the solid phase assemblage (Fig. \ref{fig:modelling_all}b) are in good agreement with their experimental counterparts within the uncertainty associated with ICP-OES measurements and the estimated surface rate constants. In contrast to its precipitation velocity, the rate of 2-line ferrihydrite dissolution is correlated to the pH and proportional to the cube of the phase mass. The rate of goethite precipitation on the other hand is sensitive to the aqueous concentration of $[\ch{Fe(OH)4^-}]^4$ and decreases exponentially with the pH (Fig. \ref{fig:rate_constants_all}).
\newpage
\clearpage
\newpage
\clearpage
For all simulations performed, the best fits have been achieved excluding a solid-solid transformation ($k_{2\text{l}\rightarrow\text{gt}} = 0 \ \si{\mole\per\square\meter\per\second}$). As displayed in Fig. \ref{fig:modelling_solid_solid}, the additional consideration of an aggregation-based transformation between 2-line ferrihydrite and goethite
predicts aqueous \ch{Fe(OH)4^-} concentrations within the confidence interval of concentration profiles predicted in its absence (Fig. \ref{fig:modelling_solid_solid}a). High solid-solid transformation rate constants in the orders of $10^{-10}$ \si{\mole\per\square\meter\per\second} result in a marginally more accurate prediction of the solid fraction at low equilibration times, but grossly over predict the rate of 2-line ferrihydrite conversion in the long term (Fig. \ref{fig:modelling_solid_solid}b). Moreover, the designated initial precipitation rate of the more soluble, amorphous precursor 2-line ferrihydrite is much larger than the rate of goethite precipitation ($\mathcal{R}_{2\text{l}}>\mathcal{R}_{\text{gt}}$). The emerging transformation mechanism, consisting of the rapid precipitation of 2-line ferrihydrite, followed by its dissolution and the slow precipitation of goethite from solution, is a multi-step process of which each step is in complete agreement with the principles of classical nucleation theory \cite{Nucleation_book} as well as Stranski's Rule \cite{blesa_stranski} and the Ostwald Step Rule \cite{ostwald1897studien}. \\
Even though the surface rate constant of goethite precipitation is $10^{23}$ to $10^{25}$ times higher than 
the rate of 2-line ferrihydrite dissolution, the effective reaction rates evaluated at the measured aqueous $n_{\ch{Fe(OH)4^-}}^{4}$ and solid $n_{2\text{l}}^3$ over time are of comparable magnitude (Fig. \ref{fig:rates_surfaces}a). At low equilibration times, the rapid decrease in $n_{\ch{Fe(OH)4^-}}$ and the corresponding reduction in the degree of supersaturation with respect to goethite appears to attenuate the high precipitation rate constant. Likewise, the specific surface area of goethite reduces to about $10\%$ of its initial value within the first 3 hours of the experiment (Fig \ref{fig:rates_surfaces}b). Over the entire timespan investigated, the rate of 2-line ferrihydrite dissolution remains below the rate of goethite precipitation, as evidenced by the strictly monotonic decrease in the aqueous $\ch{Fe(OH)4^-}$ concentration measured by ICP-OES, and can thus be considered the rate limiting step of the transformation mechanism. Fig. \ref{fig:rate_constants_all} illustrates the progression of the rates of goethite precipitation and 2-line ferrihydrite dissolution $R_{j,t}$ in \si{\mol\per\square\meter\per\second} and the evolution of their normalised specific surface area $A_{s,j,t}/A_{s,j,0}$ in \si{\square\meter\per\kilo\gram}/\si{\square\meter\per\kilo\gram}, as function of time and the pH. It can furthermore be shown that, for a specific dependence of $dn_{j,t}/dt$ on the specific surface area $A_{s,j,t}$, the evolution of $n_{j,t}$ follows the progression
\begin{equation} \label{eq:ferrihydrite_diss_tau}
	n_{j,t} = n_{0,j}\times \text{exp}(\tau(t-t_0)),
\end{equation}
where $\tau=M_{w,j}A_{s,j,0}k_j n_{j,0}^3$ in \si{\per\second}. For a more detailed account of the derivation of Equation \ref{eq:ferrihydrite_diss_tau}, the reader is referred to Supporting information, Section \ref{sec:derivation}. At constant pH and sample mass, the rate of 2-line ferrihydrite dissolution is first order with respect to $n_{j,t}$ and the time constant of dissolution $\tau$ depends entirely on the initial sample surface area. 
\newpage
\clearpage
\newpage
\clearpage
\subsection*{The precipitation of Fe(III) at neutral to alkaline pH}
\begin{figure}[!ht]
	\centering
	\includegraphics[scale=0.9]{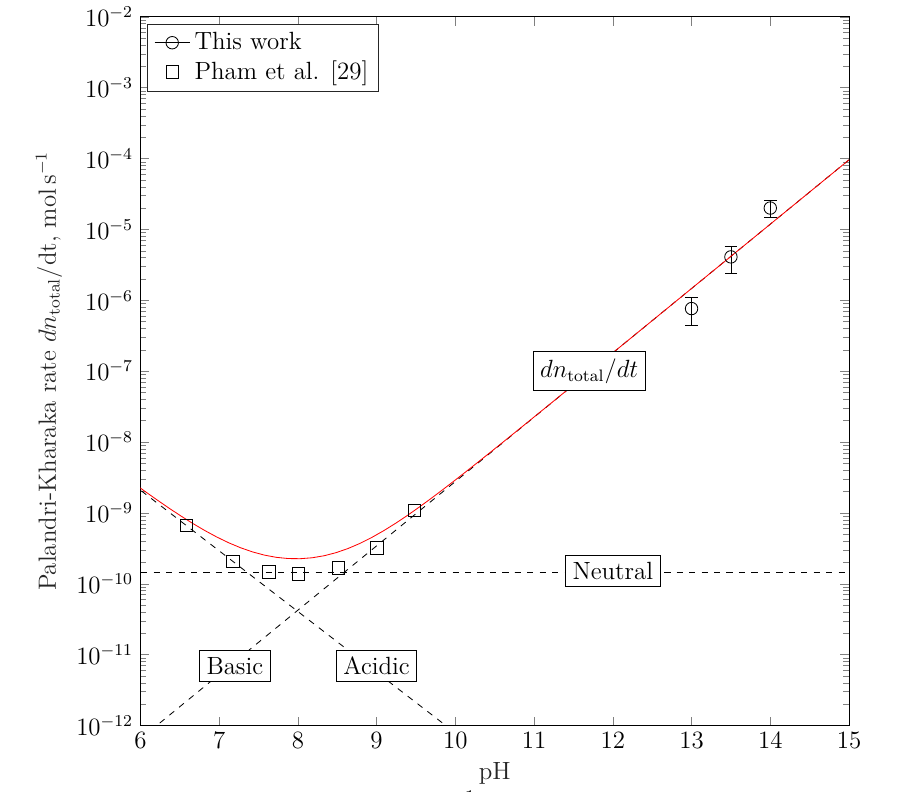}
	\caption{The estimated Palandri-Kharaka type precipitation rate of Fe(III) $dn_{\text{total}}/dt$ in \si{\mole\per\second} at neutral to alkaline pH (solid line), together with the individual mechanistic acidic, neutral and basic contributions to $dn_{\text{total}}/dt$ (dashed lines), obtained via a linear regression of the experimentally measured precipitation rates (symbols) at pH = 6.5 to 7.1, 7.6 to 8.5 and 9.4 to 14.0, respectively.}
	\label{fig:pk_rate_fits}
\end{figure}

\noindent As elucidated in the previous section, the rate of goethite precipitation is highly sensitive to the aqueous concentration of \ch{Fe(OH)4^-}. This association between precipitation velocity and the predominant aqueous Fe(III) hydrolysis product has been documented in previous studies. Pham et al. \cite{pham_precipitation} established an analogous relationship between the rate of Fe(III) precipitation and the concentration of \ch{Fe(OH)3(aq)} at pH = 6.0 to 9.5, i.e. across the predominance interval of \ch{Fe(OH)3(aq)}, and obtained an intrinsic precipitation rate constant of $k_{\ch{Fe(OH)3(aq)}}=2.0\cdot10^{7}$ \si{\liter\per\mole\per\second}. Even though Fe(III) is known to precipitate as mixtures 2-line ferrihydrite, goethite and hematite at circumneutral pH \cite{schwertmann_ph_ferri,twoline_gt_transformation}, $k_{\ch{Fe(OH)3(aq)}}$ can be compared to the transformation rates quoted in this study under the assumption that the rate of 2-line ferrihydrite dissolution is rate limiting. The observed relationship (Fig. \ref{fig:rate_constants_all}) between the rate of goethite precipitation and the concentration of \ch{Fe(OH)4^-} at pH = 13.0 to 14.0, i.e. across the predominance interval of \ch{Fe(OH)4^-}, suggests that the precipitation rate of Fe(III) follows the solubility limit of the solid Fe(III) phase stabilised. Moreover, differences in the precipitation mechanism at acidic, neutral and alkaline pH appear to be a consequence of the different Fe(III) coordination environments. The partial equilibrium model developed in this paper can hence be extended to any pH, provided the thermodynamic speciation solver includes the respective predominant aqueous Fe(III) species correlated with the precipitation velocity.
\noindent To enable better comparison between the rates goethite precipitation presented in this study and those computed by Pham et al. \cite{pham_precipitation}, various $k_{j}$ have been normalised to a surface area of 1\si{\square\meter}, multiplied by the solubility limit of 2-line ferrihydrite at pH = 6.0 to 14.0 \cite{furcas_solubility} and reformulated in terms of the \ch{H^+} activity instead of the pH. The resultant kinetic rate expression matches the notation adopted by Palandri and Kharaka \cite{palandri_model}, where the overall rate of transformation,
\begin{equation} \label{eq:pk_general_rate}
	dn_\text{total}/dt = -\bigg(k_{\text{acid}}\cdot e^{-\frac{E_a}{RT}}\cdot a_{\ch{H^+}}^p + k_{\text{neutral}}\cdot e^{-\frac{E_a}{RT}} + k_{\text{base}}\cdot e^{-\frac{E_a}{RT}}\cdot a_{\ch{H^+}}^q\bigg),
\end{equation}
consists 3 individual contributions at acidic, neutral and alkaline pH. In Equation \ref{eq:pk_general_rate}, $E_a$ stands for the activation energy in \si{\joule\per\mole}, $R$ denotes the ideal gas constant in \si{\joule\per\mole\per\kelvin} and all other parameters have their usual meanings. 
Fig. \ref{fig:pk_rate_fits} displays the fitted overall rate of Fe(III) precipitation $dn_{\text{total}}/dt$ in \si{\mole\per\second}. The kinetic parameters of each contribution to the overall rate of precipitation are obtained by segregating the experimental data into an acidic (pH = 6.5 to 7.1), near-neutral (pH = 7.6 to 8.5) and basic (pH = 9.4 to 14.0) region and then performing a piecewise linear regression on each pH interval. 
It is found that the rate constants of Fe(III) precipitation decrease from $\text{log}_{10}k=-1.49$ at acidic to $\text{log}_{10}k=-7.80$ at neutral and $\text{log}_{10}k=-15.59$ at alkaline pH. The overall transformation rate is slightly weaker correlated with $a_{\ch{H^+}}$ at acidic pH ($p=-0.857$) as opposed to the basic regime ($q=0.908$), though various acidic rate parameters may be subject to further revision due to the lack of data points at pH $ < 6.5$. Irrespective of the pH, the activation energy is estimated at $E_a = 11.7$ \si{\kilo\joule\per\mole}. It is also remarkable that the precipitation velocity at circumneutral pH is close to the dissolution rate of goethite ($\text{log}_{10}k=-7.94$), as determined by Palandri and Kharaka \cite{palandri_model}. These findings underline the crucial role of the pH dependent speciation of iron as rate limiting factor in the formation of thermodynamically stable, crystalline Fe(III) end members.
\newpage
\clearpage
\section*{Conclusion}\label{sec:Conclusion}
The partial equilibrium model presented in this study opens avenues for expanding the mechanistic understanding of the formation and transformation of iron (hydr)oxides. Such fundamental understanding is essential to solve engineering challenges in a wide variety of disciplines, ranging from the immobilisation of elements of concern in groundwater streams, soil environments, and nuclear processing facilities, the use of iron (hydr)oxides in countless industrial products, including photovoltaic and energy storage systems and pigments in the production of paints and coatings, as well as in the field of corrosion protection of iron based structures such as in the oil and gas industry, transportation infrastructure, the energy sector, and carbon storage technologies.
We illustrate the potential of this partial equilibrium model by employing it to investigate the formation and transformation of 2-line ferrihydrite to goethite. The underlying kinetic rate equations require one single fitting parameter, the rate of surface dissolution and precipitation $k_j$, to accurately predict the time-dependent solid and aqueous phase composition of the \ch{Fe}-\ch{H2O} system in disequilibrium with both solid Fe(III) phases. The precipitation of 2-line ferrihydrite is rapid. Further transformation to goethite is rate-limited by the re-dissolution of 2-line ferrihydrite, where dissolution kinetics are primarily dependent on the phase mass and specific surface area. A generalisation of the proposed trans- formation mechanism to circumneutral and mildly acidic pH suggests that the rate of Fe(III) precipitation is proportional to its own solubility limit and thus closely related to the coordination environment of the predominant aqueous Fe(III) complex.
\newpage
\clearpage
\section*{Methods}\label{sec:Methods}
The formation of 
\begin{align}
	\Gamma=\{&\ch{Fe(OH)2(s)}, \ 2\text{l}-\ch{Fe(OH)3(s)}, \ 6\text{l}-\ch{Fe(OH)3(s)}, \ \alpha-\ch{FeOOH(s)}, \ \gamma-\ch{FeOOH(s)}, \nonumber \\ &\alpha-\ch{Fe2O3(s)}\} \nonumber
\end{align}
from and in the presence of
\begin{align}
	\small \Theta =  \{ &\ch{Fe^{2+}}, \ \ch{FeOH^+}, \ \ch{Fe(OH)2(aq)}, \ \ch{Fe(OH)3^-}, \ \ch{Fe^{3+}}, \ \ch{FeOH^{2+}}, \ \ch{Fe(OH)2^+}, \nonumber \\  & \ch{Fe(OH)3(aq)}, \ \ch{Fe(OH)4^-}, \
	\ch{Fe2(OH)2^{4+}}, \ \ch{Fe3(OH)4^{5+}}, \ \ch{H2O(l)}, \ \ch{H^+}, \ \ch{OH^-}, \ \ch{Na^+}, \ \ch{Cl^-}\} \nonumber
\end{align}
can be described by a sequence of partial equilibrium steps. It is assumed that one or more $\Gamma$ is out of equilibrium with the remaining species and all $\Theta$ are in equilibrium with one another. Depending on the time-dependent evolution of the aqueous species, the rate of mineral dissolution and growth $\mathcal{R}_{j,t}$ is formulated as a function of their bulk thermodynamic quantities, the phase saturation index $\Omega_{j,t}$ and the activity of various $\Theta$ the formation of $j\in\Gamma$ is sensitive to. Analogous to the seeded growth modelling of other minerals including portlandite \cite{tardos_portlandite} and calcite \cite{dreybrodt1997precipitation}, changes to the particle geometry are considered by correcting for changes in the particle surface area upon each time step of the simulation.
\subsection*{Gibbs free energy minimisation}
Mineralogical phase equilibria and bulk compositions were determined using the Reaktoro framework \cite{reaktoro}, utilising a custom thermodynamic database of Fe(II) and Fe(III) complexes and solid phases \cite{furcas_solubility} accompanied by selected auxiliary species taken from Grenthe et al. \cite{grenthe1992chemical} and Hummel et al. \cite{hummel2002nagra}. The mineral-water interaction of $\Gamma$ can be described as
\begin{equation} \label{eq:mineral_reaction}
	\Gamma = \sum_i \nu_i a_i,
\end{equation} 
where $\nu_i$ and $a_i$ are the stoichiometric coefficient and activity of species $i\in\Theta$.  With $n^{(x)}$ being the bulk composition at equilibrium and $n^{(b)}$ representing the initial number of atoms present, the systems total Gibbs free energy
\begin{equation}
	G(n^{(x)}) = \sum_j n_j^{(x)}\mu_j \rightarrow \text{min.}
\end{equation}
is minimised subject to the molar balance
\begin{equation}
	M\cdot n^{(x)} = n^{(b)}, \qquad \forall n_j^{(x)} \geq 0.,
\end{equation}
where $M$ is the component-wise matrix of atomic balance coefficients.
\subsection*{Calculation of the activity coefficients}
We denote the chemical potential of each constituent of the aqueous electrolyte solution $i$ as
\begin{equation}
	\mu_i = \frac{\partial G(n^{(x)})}{\partial n_i} = \frac{g_i}{RT} + \text{ln}\frac{1000}{M_{w,\ch{H2O}}} + \text{ln}\frac{n_i}{n_{w}} + \text{ln}\gamma_i + 1 -\text{ln}\frac{n_{iw}}{n_w} - \frac{n_{iw}}{n_w},
\end{equation}
where $g_i$ is the partial molar Gibbs free energy in \si{\joule\per\mole} and $M_{w,\ch{H2O}} = 18.0153$ \si{\gram\per\mole} is the molecular weight of liquid water. $n_w$, $n_i$ and $n_{iw}$ represent the total number of moles of the aqueous phase, of constituent $i$ and of the water solvent $iw$ \cite{wagner2012gem,kulik2013gem}. For each species, the activity coefficient $\gamma_i$ is computed according to the extended Debye-H\"{u}ckel Equation in Truesdell-Jones form
\begin{equation} \label{eq:debye_hueckel}
	\text{log}_{10}\gamma_i = \frac{-A_\gamma z_i^2\sqrt{I}}{1 + \dot{a}B_\gamma\sqrt{I}} + b_\gamma I + \text{log}_{10}\frac{n_{iw}}{n_w},
\end{equation}
where $b_\gamma \sim 0.098$ for \ch{NaOH},  
\begin{align}
	A_\gamma &= 1.82483 \cdot 10^6 \sqrt{\varrho_0}(\varepsilon_0T)^{-3/2}, \\
	B_\gamma &= 50.2916 \cdot 10^0 \sqrt{\varrho_0}(\varepsilon_0T)^{-1/2}
\end{align}
and the effective ionic strength
\begin{equation}
	I = \frac{1}{2}\sum_j\frac{n_j}{n_{jw}}\cdot\frac{1000}{M_{w,\ch{H2O}}}z_j^2
\end{equation}
is computed from the charge of each species $z_j$ and their respective molarity, relative to 1 \si{\kilo\gram} of water \cite{helgeson1981theoretical}. Considering the density $\varrho_0$ and dielectric constant $\varepsilon_0$ of pure water at $T_r = 298.15$ K $P_r = 1$ bar, the Debye-H\"{u}ckel Limiting Law parameters are $A_\gamma \sim 0.5114$ and $B_\gamma \sim 0.3288$. 
\subsection*{Implementation of mineral-water reaction kinetics}
Dissolution and growth kinetics are incorporated in the minimisation routine via a series of partial, rather than complete equilibria, as described by Kulik and Thien \cite{kulik_thien_partial_equilibria}. The number of solid species $n^{(x)}_{j}\in\Gamma$ is changes based on its saturation index $\Omega_j$ \cite{Karpov_partial}. For 
\begin{equation} \label{eq:dualsolution}
	\Omega_j = \sum_j \text{exp}\bigg(\eta_j - \frac{g_j^\circ}{RT} - \text{ln}\gamma_j - \text{const.}\bigg),
\end{equation}  
where $\eta_j$ is the dual-solution chemical potential, $g_j^\circ$ is the standard Gibbs free energy in \si{\joule\per\mol} and $\gamma_j$ is the activity coefficient of phase $j$, the number of moles $n_j^{(x)}$ at time step $t+\Delta t$ may be computed as
\begin{equation}
	n_{j,t+\Delta t}^{(x)} = n_{j,t}^{(x)} + A_{j,t}\mathcal{R}_{j,t}\Delta t, \qquad \text{for} \qquad \text{log}_{10}\Omega_j > \epsilon_j
\end{equation}
and 
\begin{equation}
	n_{j,t+\Delta t}^{(x)} = n_{j,t}^{(x)} - A_{j,t}\mathcal{R}_{j,t}\Delta t, \qquad \text{for} \qquad \text{log}_{10}\Omega_j < \epsilon_j,
\end{equation}
where  $A_{j,t}$ is the total particle surface area, $R_{j,t}$ is the rate of phase growth in \si{\mole\per\square\meter\per\second} and $\epsilon_j$ is the stability criterion for phase $j$ at time $t$. These series simulate the stepwise precipitation from supersaturation ($\text{log}_{10}\Omega_j > 0$) and dissolution in undersaturated conditions ($\text{log}_{10}\Omega_j < 0$). The kinetic rate laws that govern changes to the bulk elemental composition of the chemical system due to the formation and dissolution of $\Gamma$ may be written as 
\begin{equation} \label{eq:partial_equilibria_rate}
	\mathcal{R}_{j,t} = k_{j}\cdot \prod_{i} a_{i,t}^{w_{i,j}}\cdot(1 - \Omega_{j,t}^{p_{j}})^{q_{j}}, \ \si{\mol\per\square\meter\per\second} \qquad \forall i \in \Theta, \ \forall j \in \Gamma, \ \forall t,
\end{equation}
where $k_{j}$ is the reaction rate constant and $\Omega_{j,t}$ denotes the dimensionless saturation index of species $j$ at time $t$. Moreover, $a_{i,t}$ represents the activity of species $i$ at time $t$ and $w_{i,j}$, $p_{j}$ and $q_{j}$ are treated as empirical parameters. For a specific molar volume of $V_{m,j}$ in \si{\cubic\meter\per\mole}, the mean orthogonal velocity of surface propagation $\mathcal{R}_{l,j,t}$ is related to the rate of phase formation $\mathcal{R}_{j,t}$,  according to 
\begin{equation} \label{eq:surface_propagation}
	\mathcal{R}_{l,j,t} = V_{m,j}\times \mathcal{R}_{j,t}.
\end{equation} 
In these Palandri-Kharaka type reaction rate expressions \cite{palandri_model}, the saturation index is evaluated directly from the dual-solution chemical potential of phase $j$, as displayed in Equation \ref{eq:dualsolution}.
\subsection*{Surface area and morphology correction}
\begin{figure}
	\centering
	\includegraphics[scale=2.7]{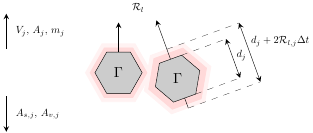}
	\caption{Schematic illustration of the growth process of mineral particles at the the mean orthogonal velocity of surface propagation $\mathcal{R}_{l,j}$. The resultant increase of the particle diameter and thus particle volume $V_j$, surface area $A_j$ and mass $m_j$ causes a reduction in specific surface area $A_{s,j}$ and surface area per unit volume $A_{v,j}$.}
	\label{fig:mineral_growth}
\end{figure}

Changes to the mineral surface area $A_j$ are incorporated into the molar balance of each phase $j\in\Gamma$ by two different models part of the TKinMet library of the geochemical modelling package GEM-Selektor \cite{kulik2013gem,kulik_thien_partial_equilibria}. Consider the growth of $\Gamma$, as schematically illustrated in Fig. \ref{fig:mineral_growth}.
The increase of particle diameter $d_j$ at time $t$ to $d_j + 2\mathcal{R}_{l,j}\Delta t$ at time $t + \Delta t$ at the mean orthogonal rate of surface propagation $\mathcal{R}_{l,j}$ increases the particle volume $V_j$, surface area $A_j$ and mass $m_j$, whilst the specific surface area $A_{s,j}=A_j/m_j$ and the area per unit volume $A_{v,j}=A_j/V_j$ are expected to reduce. This reduction can be computed from the initial specific surface area $\text{A}_{s,j,0}$ by the simple cubic root correction 
\begin{equation} \label{eq:mass_cubic_root}
	A_{s,j,t} = A_{s,j,0} \cdot \bigg( \frac{n_{j,0}}{n_{j,t}} \bigg)^{1/3},
\end{equation} 
where $n_{j,0}$ and $n_{j,t}$ represent the initial and final number of moles of phase $j$ \cite{RN241,thien_area_correction}. The expected reduction to the surface area per unit volume can alternatively be described by 
\begin{equation} \label{eq:areapervolume}
	A_{v,j,t} = A_{v,j,t-\Delta t}\times\frac{\psi_{j,t-\Delta t}}{\psi_{j,t}}\times\frac{d_{j,t-\Delta t}}{d_{j,t-\Delta t} + 2R_{l,j,t}},
\end{equation}
where all parameters have their usual meaning. The shape factor $\psi_j$ in Equation \ref{eq:areapervolume} is equivalent to the sphericity coefficient, as described by Wadell \cite{wadell_sphericity}:
\begin{equation}
	\psi_j=\pi^{1/3}\frac{(6V_j)^{1/3}}{A_j}=\frac{6V_j}{d_jA_j},
\end{equation}
where $V_j$ and $A_j$ are the mean particle volume and surface area and $d_j=6/(\psi_jA_j)$ is the estimated particle diameter. Further, changes in morphology upon dissolution and growth are accounted for by the shape factor function, expressed as a formal power series 
\begin{equation}
	\psi_j(t) = \psi_{j,0} + \psi_{j,1}u + \psi_{j,2}u^2 + \dots + \psi_{j,n}u^n, 
\end{equation}
of the phase saturation index $u = \text{log}_{10}\Omega_{j,t}$ \cite{thien_area_correction}.

\backmatter

\bmhead{Supplementary information}

Supplementary information is available for this publication under the heading Extended Data.

\section*{Declarations}
\subsection*{Funding}
The authors are grateful to the European Research Council (ERC) for the financial support provided for Fabio Enrico Furcas under the European Union Horizon 2020 research and innovation program (grant agreement no. 848794). 
\subsection*{Conflict of interest}
The authors declare no competing interests.
\subsection*{Availability of data and materials}
The data that support the findings of this study are available within the article and its Supplementary Information.
\subsection*{Ethics approval}
Not applicable
\subsection*{Consent to participate}
Not applicable
\subsection*{Consent for publication}
Not applicable
\subsection*{Authors' contributions}
Fabio E. Furcas, Shishir Mundra, Barbara Lothenbach and Ueli M. Angst conceived the overall study; all authors contributed to the study design, the analysis and interpretation of the results. Fabio E. Furcas wrote the main draft of the manuscript, to which all authors contributed. Ueli M. Angst was the main supervisor of the project. All authors read and approved the final manuscript.
\newpage
\clearpage
\begin{appendices}
\section{Extended Data}
\subsection{Derivation of the rate of 2-line ferrihydrite dissolution} \label{sec:derivation}
The molar rate of 2-line ferrihydrite dissolution
\mathleft
\begin{align} 
	dn_{j,t}/dt &= A_{j,t}\mathcal{R}_{j,t}  \tagaddtext{$(\si{\square\meter})\cdot(\si{\mole\per\square\meter\per\second})$} \nonumber \\
	&= A_{s,j,t} M_{w,j} n_{j,t} k_j n_{j,t}^3  \tagaddtext{$(\si{\square\meter})\cdot(\si{\mole\per\square\meter\per\second})$}  \nonumber \\
	&= A_{s,j,t} M_{w,j} k_j n_{j,t}^4 \tagaddtext{$(\si{\square\meter\per\gram})\cdot(\si{\gram\per\mol})\cdot(\si{\per\square\mole\per\square\meter\per\second})\cdot(\si{\mole}^4)$} \label{eq:ferrihydrite_dissolution_molar}
\end{align}
can be expressed in terms of the initial specific surface area, by using the cubic root correction, as displayed in Equation \ref{eq:mass_cubic_root}, or any other general correction formula
\begin{equation} \label{eq:mass_cubic_root_general}
	A_{s,j,t} = A_{s,j,0} \cdot \bigg( \frac{n_{j,0}}{n_{j,t}} \bigg)^{\alpha} \tagaddtext{(\si{\square\meter})}.
\end{equation}
Substituting $A_{s,j,t}$ in Equation \ref{eq:ferrihydrite_dissolution_molar} with Equation \ref{eq:mass_cubic_root_general} yields
\begin{align}
	dn_{j,t}/dt &= A_{s,j,0} n_{j,0}^\alpha n_{j,t}^{-\alpha}  M_{w,j} k_j n_{j,t}^{4}
	= A_{s,j,0} n_{j,0}^\alpha M_{w,j} k_j n_{j,t}^{4-\alpha}.  \nonumber \\ \tagaddtext{$(\si{\square\meter\per\gram})\cdot(\si{\mole}^\alpha)\cdot(\si{\gram\per\mol})\cdot(\si{\per\square\mole\per\square\meter\per\second})\cdot(\si{\mole}^{4-\alpha})$} \label{eq:ferrihydrite_dissolution_alpha}
\end{align}
Integrating from $n_{j,0}$ to $n_{j,t}$ and $t_0$ to $t$,
\begin{equation}
	\int_{n_{j,0}}^{n_{j,t}}dn_{j,t}n_{j,t}^{\alpha-4} =  A_{s,j,0} n_{j,0}^\alpha M_{w,j} k_j\int_{t_0}^t dt, \tagaddtext{(\si{\mol})}
\end{equation}
the number of moles of ferrihydrite $n_{j,t}$ decay exponentially for $\alpha=3$,
\begin{equation}
	n_{j,t} = n_{j,0}\times \text{exp}(A_{s,j,0}M_{w,j}k_jn_{0,j}^3(t-t_0)) = n_{j,0}\times\text{exp}(\tau(t-t_0)), \tagaddtext{(\si{\mol})}
\end{equation}
where $\tau = A_{s,j,0}M_{w,j}k_jn_{0,j}^3$ in \si{\per\second} is the time constant of dissolution.
For $\alpha = 4$, $n_{j,t}$ reduces linearly according to
\begin{equation} 
	n_{j,t} = n_{j,0} + A_{s,j,0}M_{w,j}k_jn_{0,j}^4(t-t_0) = \tau(t-t_0), \tagaddtext{(\si{\mol})} \label{eq:tau_alpha_4}
\end{equation}
and for $\alpha \neq 3,4$, the progression of $n_{j,t}$ is described by
\begin{equation}
	n_{j,t} = \bigg(n_{j,0}^{\alpha-3} + (\alpha-3)A_{s,j,0}M_{w,j}k_jn_{0,j}^\alpha(t-t_0)\bigg)^{1/(\alpha-3)}. \tagaddtext{(\si{\mol})} \label{eq:tau_alpha_no34}
\end{equation}
Note that $\tau$ in Equations A6 and A7, $\tau$ is not a real time constant and has units of $\si{\mole}^{\alpha-3}\si{\per\second}$.
\newpage
\clearpage
\subsection{List of symbols and notations}
\begin{table}[!ht]
	\centering
	\setlength\extrarowheight{5pt}
	\caption{List of symbols and notations used in this paper. In addition to the parameter-specific subscripts listed in this table, indices $i,j$ refer to the chemical species $\Gamma$, $\Theta$ and index $t$ denotes time.}
	\label{tab:listofsymbols}
		\begin{tabular*}{\textwidth}{l l l l}
			\hline\hline
			Symbol & Description & 	Symbol & Description  \\
			\hline
			$A$ & Particle surface area, \si{\square\meter} & $n$ & Number of moles, \si{\mole}\\
			$A_s$ & Specific surface area (SSA), \si{\square\meter\per\kilo\gram} & $\nu$ & Stoichiometric coefficient, / \\
			$A_v$ & Surface area per unit volume, \si{\square\meter\per\cubic\meter} & 		 $\Omega$ & Saturation index, / \\
			$a$ & Chemical activity, / & 	$P$ & Pressure, \si{\bar}\\
			$d$ & Particle diameter, \si{\meter} & 	$p$ & Empirical parameter, / \\
			$E_a$ & Activation energy, \si{\joule\per\mole} & $q$ & Empirical parameter, / \\
			$G$ & Total Gibbs free energy, \si{\joule} & $\psi$ & Wadell sphericity \cite{wadell_sphericity} , / \\
			$g$ & Partial molar Gibbs free energy, \si{\joule\per\mole} & 	$\mathcal{R}$ & Rate of phase formation, \si{\mol\per\square\meter\per\second} \\
			$g^\circ$ & Standard molar Gibbs free energy, \si{\joule\per\mole} & 	$\mathcal{R}_l$ & Mean orthogonal rate of surface propagation, \si{\meter\per\second}\\
			$\gamma$ & Activity coefficient, / & $R$ & Ideal gas constant, 8.314 \si{\joule\per\mole\per\kelvin} \\
			$\nu$ & Dual-solution chemical potential, / & $\varrho$ & Density, \si{\kilo\gram\per\cubic\meter} \\
			$\epsilon$ & Phase stability criterion, / & 	$T$ & Temperature, \si{\kelvin} \\
			$\varepsilon$ & Dielectric constant, / &	$u$ & Decadic logarithm of the phase saturation index, / \\
			$I$ & Effective ionic strength, \si{\mole\per\kilo\gram} & 	$V$ & Particle volume, \si{\cubic\meter}\\
			$k$ & Reaction rate constant, \si{\mole\per\square\meter\per\second} & 	$V_m$ & Specific molar volume, \si{\cubic\meter\per\mole}\\
			$M_w$ & Molecular weight, \si{\gram\per\mole} & 	$w$ & Reaction order term, / \\
			$m$ & Particle mass, \si{\kilo\gram} & 		$z$ & Formal charge, / \\
			$\mu$ & Chemical potential, /  & $[ \ \ \ ]$ & Concentration, \si{\mole\per\liter}\\
			\hline\hline 
		\end{tabular*}
\end{table}%
\subsection{Additional plots}
\begin{figure}[!ht]
	\centering
	\begin{subfigure}[b]{\textwidth}
		\centering
		\includegraphics[scale=0.9]{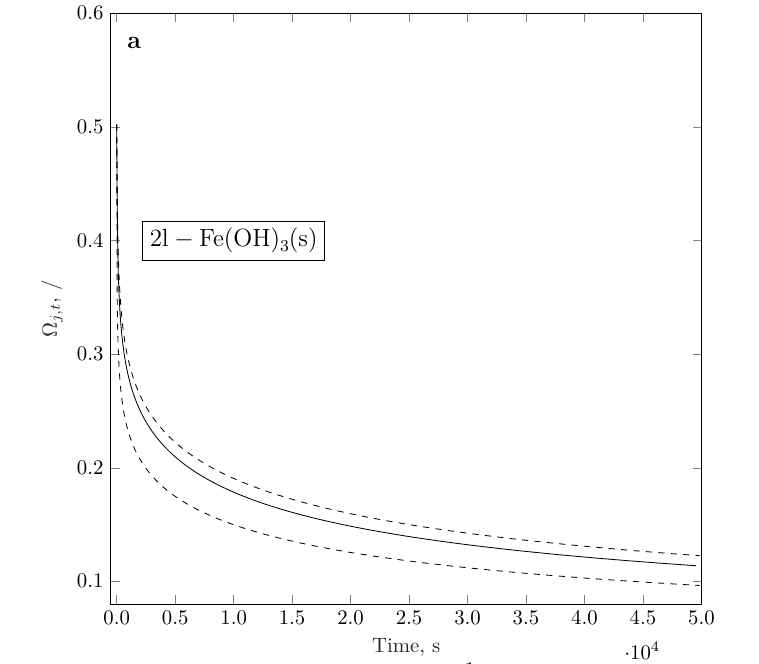}
	\end{subfigure}%
	\vskip\baselineskip
	\begin{subfigure}[b]{\textwidth}
		\centering
		\includegraphics[scale=0.9]{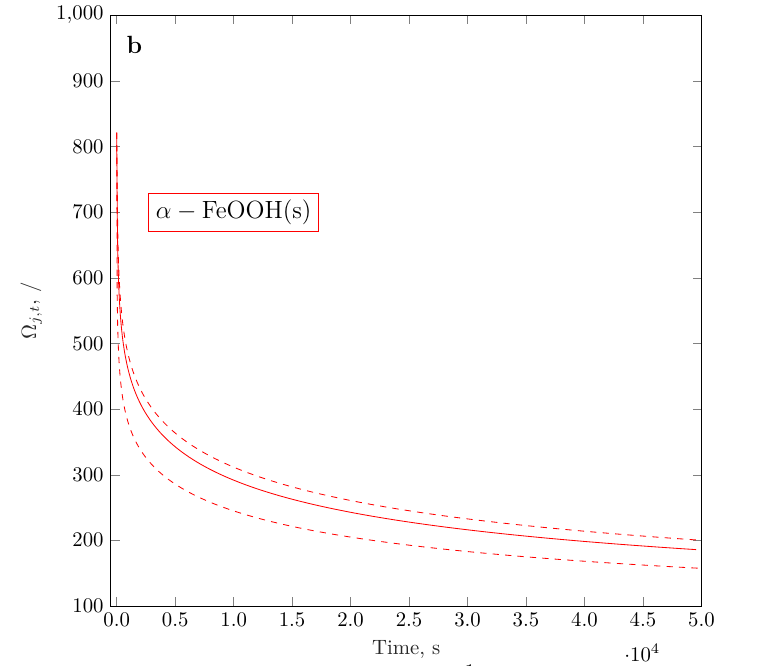}
	\end{subfigure}
	\caption{Saturation indices of species 2-line ferrihydrite (Fig. \ref{fig:saturation_indices}a) and goethite (Fig. \ref{fig:saturation_indices}b) at pH = 14.0 over time. Dashed lines represent the confidence interval of predicted saturation indices, corresponding to the upper and lower limit of reaction rate constants displayed in Fig. \ref{fig:rate_constants_all}.}
	\label{fig:saturation_indices}
\end{figure}
\newpage
\clearpage


\end{appendices}


\bibliography{sn-bibliography}

\end{document}